\documentclass[aps,showpacs,twocolumn,pra,superscriptaddress]{revtex4-1}

\usepackage[tbtags]{amsmath}
\usepackage{amssymb}
\usepackage{amsfonts}
\usepackage{graphicx,color}
\usepackage{bmpsize}
\usepackage{mathtools}
\usepackage[colorlinks=true,citecolor=blue,urlcolor=blue]{hyperref}

\hypersetup{colorlinks,linkcolor=blue,filecolor=blue,urlcolor=blue,citecolor=blue,}
\usepackage{multirow}

\usepackage{float}

\usepackage{appendix}
\begin{document}
	\title{Sharing Quantum Steering via Standard Projective Measurements}

	\author{Shufen Dong}
	\affiliation{Key Laboratory of Low-Dimension Quantum Structures and Quantum Control of Ministry of Education, Synergetic Innovation Center for Quantum Effects and Applications, Xiangjiang-Laboratory and Department of Physics, Hunan Normal University, Changsha 410081, China}
	\author{Zinuo Cai}
	\affiliation{Key Laboratory of Low-Dimension Quantum Structures and Quantum Control of Ministry of Education, Synergetic Innovation Center for Quantum Effects and Applications, Xiangjiang-Laboratory and Department of Physics, Hunan Normal University, Changsha 410081, China}
	
	\author{Chunfeng Wu}\thanks{Corresponding author: chunfeng\_wu@sutd.edu.sg}
	\affiliation{Science, Mathematics and Technology, Singapore University of Technology and Design, 8 Somapah Road, Singapore 487372, Singapore}
	
	\author{Changliang Ren}\thanks{Corresponding author: renchangliang@hunnu.edu.cn}
	\affiliation{Key Laboratory of Low-Dimension Quantum Structures and Quantum Control of Ministry of Education, Synergetic Innovation Center for Quantum Effects and Applications, Xiangjiang-Laboratory and Department of Physics, Hunan Normal University, Changsha 410081, China}

	\begin{abstract}
		We propose a scheme for the sharing of quantum steering among three observers, Alice, Bob, and Charlie using standard projective measurements. We show that in the unilateral sequential scenario, Alice can steer Bob's and Charlie's states and conversely, Bob and Charlie can steer Alice's state. Unlike the quantum steering sharing achieved through weak measurements, we use the standard projective measurements to enable quantum steering sharing. Quantum steering is demonstrated by the violations of the linear steering inequality among different observer combinations. We find that Alice can simultaneously steer both Bob's and Charlie's states, and Bob and Charlie can simultaneously steer Alice's state, regardless of whether they are in maximally entangled states or partially entangled states. The maximum double violation of the linear steering inequalities obtained from partially entangled states can be greater in some cases than that obtained from maximally entangled states when randomly combining the case of two projective measurements and the case of two identity measurements. Additionally, we verify hybrid quantum correlation sharing through the double violation of the Clauser-Horne-Shimony-Holt (CHSH) inequality and the linear steering inequality. Our results provide a new perspective for the study of quantum steering and may lead to applications in quantum random access code, randomness certification, and self-testing process.
	\end{abstract}

	%================================================================================
	
	\maketitle
	
	\section{Introduction}
	The exploration of quantum correlation traces back to 1935 when Einstein, Podolsky, and Rosen (EPR) challenged the completeness of quantum mechanics in the EPR paradox \cite{Einstein.PhysRev.47.777_1935}.  Over the years, three types of non-classical correlations have emerged from the EPR paradox subsequently, including quantum entanglement 
	\cite{Werner.PhysRevA.40.4277_1989,Horodecki.Phys.Lett.A.223.1_1996,Horodecki.RevModPhys.81.865_2009,Gühne.PhysRep.474.1_2009,Lami.PhysRevLett.117.220502_2016}, EPR steering \cite{schrödingerProc.CambridgePhilos.Soc.31.555_1935,schrödinger.Proc.CambridgePhilos.Soc.32.446_1936,Wiseman.PhysRevLett.98.140402_2007,Reid.PhysRevA.40.913_1989,Cavalcanti.PhysRevA.80.032112_2009,Reid.PhysRevA.88.062108_2013,Cavalcanti.J.Opt.Soc.Am.B.32.A74_2015}, and Bell non-locality \cite{Bell.Physics.1.195_1964,Clauser.PhysRevLett.23.880_1969,Aspect.PhysRevLett.49.91_1982,Toner.Proc.R.Soc.A465.59_2009,Brunner.RevModPhys.86.419_2014}. On one hand, the quantum correlations have deepened our understanding of fundamental concepts in quantum mechanics. On the other hand, the quantum correlations have demonstrated their promising capability in quantum computing \cite{Zhou.PhysRevX.10.041038_2020,Fitzsimons.PhysRevLett.120.040501_2018,Zhong.PhysRevLett.127.180502_2021,Madsen.Nature.606.75_2022,Haug.PRXQuantum.4.010301_2023}, quantum communications \cite{Gisin.Nat.Photonics.1.165_2007,Pirandola.Nat.Commun.8.15043_2017,Bouwmeester.Nature.390.575_1997,Del.Santo.PhysRevLett.120.060503_2018,Luo.Yi-Han.PhysRevLett.123.070505_2019,Hu.XiaoMin.PhysRevLett.126.010503_2021}, quantum metrology \cite{Giovannetti.PhysRevLett.96.010401_2006,Szczykulska.ADV.PHYS.X.1.621_2016,Koczor.NewJ.Phys.22.083038_2020,Barbieri.3.010202_2022,Lee.Phys.Rev.Res.5.013103_2023}, and other pioneering fields \cite{Maccone.Lorenzo.PhysRevLett.124.200503_2020,Georgescu.RevModPhys.86.153_2014}.
	
	Among the quantum correlations, Bell nonlocality and quantum steering are commonly detected by the violation of certain inequalities via local measurements performed by individual observers on a shared quantum state. Specially, optimal quantum correlations have been extensively investigated in the literature through the largest violation of inequalities. The measurements utilized to reveal optimal quantum correlations are generally the so-called strong measurements \cite{Neumann.1955}, that correspond to projective measurements. Such measurements maximally perturb the state, rendering it separable after the measurements. In this scenario, quantum correlations cannot be collectively shared among observers who perform sequential measurements.
	
	In 2015, Silva \emph{et al.} explored an alternative scenario in which a single observer (named as Alice) and multiple observers (named as Bobs) share an entangled pair, with the intermediate Bobs performing weak measurements \cite{Silva.PhysRevLett.114.250401_2015}. It was shown that the Bell nonlocality of a pair of qubits may be shared among more than two observers and this result was experimentally demonstrated in Refs. \cite{Matteo.Schiavon.Quantum.Sci.Technol.2.015010_2017} two years later. This intriguing finding not only broadened the understanding on the properties of quantum correlations, but also stimulated further explorations about the recycling of Bell nonlocality \cite{Mal.Shiladitya.Mathematics.4.48_2016,Hu.MengJun.npj.Quantum.Inf.4.63_2018,Ren.Changliang.PhysRevA.100.052121_2019,Das.Debarshi.PhysRevA.99.022305_2019,Feng.Tianfeng.PhysRevA.102.032220_2020,Foletto.Giulio.Phys.Rev.Appl.13.044008_2020,Brown.Peter.PhysRevLett.125.090401_2020,Cheng.Shuming.PhysRevA.104.L060201_2021,Ren.Changliang.PhysRevA.105.052221_2022,Zhang.Tinggui.PhysRevA.103.032216_2021,Zinuo.Cai.arXiv_2024}.  Meanwhile, similar investigations have been extended to the sharing of other types of quantum resources, like quantum steering \cite{Yao.Dan.PhysRevA.103.052207_2021,Sasmal.Souradeep.PhysRevA.98.012305_2018,Yeon.Ho.Choi.Optica.7.675_2020,ShenoyH.PhysRevA.99.022317_2019,Zhu.Jie.PhysRevA.105.032211_2022,Gupta.Shashank.PhysRevA.103.022421_2021,Liu.TongJun.Opt.Express.30.41196_2022,Lv.QiaoQiao.J.Phys.A.Math.Theor.56.325301_2023,Chen.Yuyu.Chin.Phys.B.32.040309_2023}, quantum contextuality \cite{Kumari.Asmita.PhysRevA.100.062130_2019,Anwer.Hammad.Quantum.5.551_2021,Kumari.PhysRevA.107.012615_2023}, quantum entanglement \cite{Bera.Anindita.PhysRevA.98.062304_2018,Srivastava.Chirag.PhysRevA.103.032408_2021,Pandit.PhysRevA.106.032419_2022,MaoSheng.Li.arXiv_2023},  and network nonlocal sharing \cite{Hou.Wenlin.PhysRevA.105.042436_2022,Mahato.ShyamSundar.PhysRevA.106.042218_2022,Cai.Zinuo.arXiv_2022,Mahato.PhysRevA.106.042218_2022,Kumar.Rahul.Quantum.Stud.Math.Found.10.353_2023}, etc. From the perspective of practical application, the recycling or sharing of different quantum correlations has been shown to play vital roles in performing various quantum tasks, such as quantum random access code \cite{Mohan.New.J.Phys.21.083034_2019,Anwer.Hammad.PhysRevLett.125.080403_2020,Wei.Shihui.New.J.Phys.23.053014_2021,Foletto.Giulio.PhysRevResearch.2.033205_2020,Das.Debarshi.PhysRevA.104.L060602_2021}, randomness certification \cite{Curchod.F.J.PhysRevA.95.020102_2017,Foletto.PhysRevA.103.062206_2021,Bowles.Quantum.4.344_2020,Coyle.cryptography.3.27_2019,Florian.arXiv_2018}, and self-testing process \cite{Miklin.Nikolai.PhysRevResearch.2.033014_2020,Roy.Prabuddha.New.J.Phys.25.013040_2023}.
	
	In the references, most of the results have been achieved with a common condition that the recycling of different quantum correlations is through unsharp measurements. It is also interesting to investigate the recycling or sharing of quantum correlations without the constraint of unsharp measurements. Very recently, Stenlongo \emph{et al.} proposed a protocol for distributing Bell nonlocality among Alice and a series of Bobs. In their approach, Alice and each Bob stochastically employ three distinct types of projective measurement strategies \cite{Steffinlongo.PhysRevLett.129.230402_2022}. The experimental demonstration of the recycling of nonlocal resources with projective measurements has then been realized by Xiao \emph{et. al} \cite{Ya.Xiao.arXiv_2022}. Subsequently, Zhang \emph{et al.} achieved the nonlocal sharing of arbitrary high-dimensional pure bipartite states using projective measurements \cite{ZhangTinggui.PhysRevA.109.022419_2024}. Here, we investigate whether quantum steering can also be recycled using only standard projective measurements. Specifically, we consider the unilateral sequential scenario in which a two-qubit entangled state is shared among Alice, Bob, and Charlie. This is achieved by sending one qubit to Alice and another qubit to Bob and Charlie sequentially. We first demonstrate that Alice can steer Bob's and Charlie's states simultaneously through the standard projective measurements, as well as Bob and Charlie can steer Alice's state simultaneously given a maximally entangled state. We then extend this analysis to partially entangled states, and still observe the steering sharing based on the standard projective measurements. In the latter case, the double violation is greater in some cases than that for maximal entangled states as expected when randomly combining the case of two projective measurements and the case of two identity measurements. Finally, we discuss the hybrid quantum correlation sharing by the double violation of the CHSH inequality (between Alice and Bob) and the linear steering inequality (between Alice and Charlie).

	The paper is organized as follows. In Sec. \hyperref[1]{\uppercase\expandafter{\romannumeral2}}, we explain the unilateral sequential model and criteria. In Sec. \hyperref[2]{\uppercase\expandafter{\romannumeral3-A}}, we explore the ability of Alice to steer Bob's and Charlie's states, while in Sec. \hyperref[3]{\uppercase\expandafter{\romannumeral3-B}}, we investigate hybrid quantum correlation sharing. Conversely, it is shown that Bob and Charlie can also steer Alice's state in Sec. \hyperref[4]{\uppercase\expandafter{\romannumeral4-A}}, and hybrid quantum correlation sharing is illustrated in  Sec. \hyperref[5]{\uppercase\expandafter{\romannumeral4-B}}. The last section is for conclusion.
	
	\section{The scenario of EPR-steering sharing} \label{1}
\begin{figure}[htbp]
		\vspace{-2mm} 
		\begin{center}
			\includegraphics[width=\columnwidth]{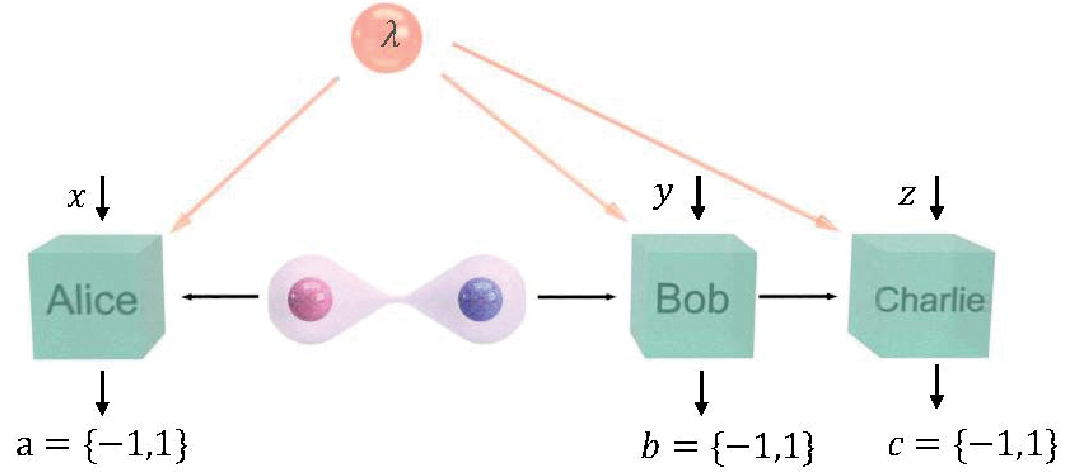} 
			
			\caption{The unilateral sequential scenario: a pair of entangled qubits are distributed to Alice, Bob, and Charlie. One qubit is sent to Alice and the other one is sent sequentially to Bob and Charlie. Before each round of the experiment, all observers need to share relevant classical data strings $\lambda$, which are random variables whose values follow some probability distribution $\{p\left(\lambda\right)\}_{\lambda}$. } \label{figure 1}
		\end{center}
		\vspace{-5mm}  
	\end{figure}

	We mainly study the simplest unilateral sequential scenario which is similar to that in \cite{Steffinlongo.PhysRevLett.129.230402_2022}, as shown in Fig. \ref{figure 1}. A two-qubit entangled source $\rho$ distributes qubits to Alice and Bob. They measure two different dichotomic observables $\hat{A}_x$ and $\hat{B}_y$ with binary outcome $a\in\{-1,1\}$ and $b\in\{-1,1\}$, where $x\in\{0,1\}$ and $y\in\{0,1\}$. Then Bob performs unitary operation $\hat{U}_y$ on his post-measurement qubit and sends it to Charlie. Charlie performs the measurement $\hat{C}_z$ with the corresponding outcome $c\in\{-1,1\}$, where $z\in\{0,1\}$. All observers perform independent projective measurements, and the choice of measurement is unbiased. After the measurements of all observers are completed, we can explore the quantum steering of two different pairs based on the combination of observers (Alice-Bob and Alice-Charlie). The quantum steering can be verified by the violation of the linear steering inequality \cite{Cavalcanti.PhysRevA.80.032112_2009}. The linear steering inequalities for Alice-Bob and Alice-Charlie are given as
	\begin{align}
		S_{L_1}=\frac{1}{\sqrt{2}}\left|\sum_{i=0}^{1}\left\langle A_i\otimes B_i\right\rangle\right|\le1\label{a},\\
		S_{L_2}=\frac{1}{\sqrt{2}}\left|\sum_{i=0}^{1}\left\langle A_i\otimes C_i\right\rangle\right|\le1\label{b},
	\end{align}
	where $S_{L_1}$ and $S_{L_2}$ represent the linear steering parameters of Alice-Bob and Alice-Charlie respectively, and $\left\langle A_i\otimes B_i\right\rangle=Tr[\hat{A}_i\otimes \hat{B}_i.\rho]$, $\left\langle A_i\otimes C_i\right\rangle=Tr[\hat{A}_i\otimes \hat{C}_i.\rho_{AC}]$. As Charlie's measurement is not dependent on Bob's measurement choices and outcomes, the state shared between Alice and Charlie is given by $\rho_{AC}=\frac{1}{2}\sum_{b,y}\left(I\otimes \sqrt{\hat{E}_{b|y}}\right).\rho.\left(I\otimes \sqrt{\hat{E}_{b|y}}\right)^{\dagger}$, where $\hat{E}_{b|y}$ is Kraus operator and $\hat{E}_{b|y}=\hat{U}_{b|y}\hat{B}_{b|y}$. If the Alice-Bob pair and the Alice-Charlie pair can simultaneously violate the linear steering inequalities, quantum steering sharing can be achieved. 
	
	In this scenario, if Bob performs two possible projective measurements, it will destroy the entanglement of state $\rho$, turning it into a separable state, thereby preventing the sharing of quantum state resource information. However, if Bob performs either two identity measurements or a projective measurement and an identity measurement, meaning that one or both measurements fail, the post-measurement state will remain entangled, enabling the sharing of quantum state resource information with Charlie. We adopt a measurement protocol proposed in \cite{Steffinlongo.PhysRevLett.129.230402_2022}, which combines the case of two projective measurements for Bob with $\lambda=1$, the case of two identity measurements for Bob with $\lambda=2$, and the case of a projective measurement and an identity measurement for Bob with $\lambda=3$ with some random probability distribution $\{p\left(\lambda\right)\}_{\lambda=1}^3$. This measurement protocol can achieve nonlocal sharing, which has also been experimentally demonstrated in \cite{Ya.Xiao.arXiv_2022}. Therefore, we explore quantum steering sharing by employing this measurement protocol. At this time, the linear steering inequality between Alice and Bob, as well as Alice and Charlie, can be described as 
	\begin{align}
		S_1&=\sum_{\lambda=1}^3p\left(\lambda\right)S_{L_1}^{(\lambda)}\le1, \nonumber\\
		S_2&=\sum_{\lambda=1}^3p\left(\lambda\right)S_{L_2}^{(\lambda)}\label{c}\le1,
	\end{align}
	where (\ref{a}) and (\ref{b}) are also applicable to $S_{L_1}^{(\lambda)}$ and $S_{L_2}^{(\lambda)}$ respectively. We will discuss two types of quantum steering sharing in the following: one is Alice to steer the states of Bob and Charlie, and the other is Bob and Charlie to steer Alice's state. Besides, we also discuss the hybrid quantum correlation sharing composed of quantum steering and nonlocality. 
	%where $\lambda\in\{1,2,3\}$ denotes that all observers need to share correlated strings of classical data and they are used to determine the measurement cases.
	
	\section{Alice to steer Bob's and Charlie's states} 
	In this section, we study the ability of Alice to steer the states of Bob and Charlie. %\textcolor{red}{We specify the measurement directions of the steered parties to be vertical to observe the steering phenomenon more obviously.}
	Without loss of generality, Alice's two observables can be defined as
	$\hat{A}_x=\cos\theta'\sigma_1+(-1)^x\sin\theta'\sigma_3$, where $\{\sigma_1, \sigma_2, \sigma_3\}$ are the Pauli matrices. Quantum steering sharing and hybrid quantum correlation sharing are then illustrated by exploring the two linear steering inequalities and the CHSH inequality together with one linear steering inequality in the following.
	\subsection{The case of two linear steering inequalities} \label{2}
	
	Considering a maximally entangled state $\rho=\left|\psi\right\rangle\left\langle\psi\right|,\left|\psi\right\rangle=\frac{1}{\sqrt{2}}\left(\left|00\right\rangle+\left|11\right\rangle\right)$, the optimal measurement settings for the above-mentioned three measurement cases are shown below. 
	
	Case 1. When $\lambda=1$, both measurements for Bob are basis projections. Alice's measurement choices are $\hat{A}_0^{(1)}=\cos\theta\sigma_1+\sin\theta\sigma_3$ and $\hat{A}_1^{(1)}=\cos\theta\sigma_1-\sin\theta\sigma_3$. Bob's measurement choices are $\hat{B}_0^{(1)}=\frac{\sqrt{2}}{2}\sigma_1-\frac{\sqrt{2}}{2}\sigma_3$ and $\hat{B}_1^{(1)}=\frac{\sqrt{2}}{2}\sigma_1+\frac{\sqrt{2}}{2}\sigma_3$, and the corresponding unitary operations are $\hat{U}_0^{(1)}=I, \hat{U}_1^{(1)}=e^{-\frac{\pi}{4}i\sigma_2}$. Charlie's measurement choices are $\hat{C}_0^{(1)}=\hat{C}_1^{(1)}=\frac{\sqrt{2}}{2}\sigma_1-\frac{\sqrt{2}}{2}\sigma_3$. $I$ is the identity matrix. According to (\ref{a}) and (\ref{b}), we obtain $S_{L_1}^{(1)}=\left|\cos\theta-\sin\theta\right|$, $S_{L_2}^{(1)}=\left|\cos\theta\right|$. To visually observe the relationship between $S_{L_1}^{(1)}$ and $S_{L_1}^{(2)}$, we define a trade-off. That is, given a specific value of $S_{L_1}^{(1)}$, there exists a possible corresponding value of $S_{L_1}^{(2)}$. So the trade-off is $S_{L_2}^{(1)}=\frac{1}{2}\left(S_{L_1}^{(1)}+\sqrt{2-\left(S_{L_1}^{(1)}\right)^2}\right)$ in the range of $S_{L_1}^{(1)}\in\left[1,\sqrt{2}\right]$ and we easily obtain the classic trade-off $S_{L_2}^{(1)}=1$ in the range of $S_{L_1}^{(1)}\in\left[0,1\right]$.
	
	Case 2. When $\lambda=2$, both measurements for Bob are identity measurements. Alice's measurement choices are $\hat{A}_0^{(2)}=\frac{\sqrt{2}}{2}\sigma_1+\frac{\sqrt{2}}{2}\sigma_3$ and $\hat{A}_1^{(2)}=\frac{\sqrt{2}}{2}\sigma_1-\frac{\sqrt{2}}{2}\sigma_3$. Bob's measurement choices are $\hat{B}_0^{(2)}=\hat{B}_1^{(2)}=I$, and the corresponding unitary operations are $\hat{U}_y^{(2)}=I$. Charlie's measurement choices are $\hat{C}_0^{(2)}=\frac{\sqrt{2}}{2}\sigma_1+\frac{\sqrt{2}}{2}\sigma_3$ and $\hat{C}_1^{(2)}=\frac{\sqrt{2}}{2}\sigma_1-\frac{\sqrt{2}}{2}\sigma_3$. We obtain $S_{L1}^{(2)}=0$, $S_{L2}^{(2)}=\sqrt{2}$, where $S_{L2}^{(2)}$ reaches the quantum bound of the linear steering inequality.
	
	Case 3. When $\lambda=3$, one measurement is a basis projection and the other is an identity measurement for Bob. Alice's measurement choices are $\hat{A}_0^{(3)}=\cos\delta\sigma_1+\sin\delta\sigma_3$ and $\hat{A}_1^{(3)}=\cos\delta\sigma_1-\sin\delta\sigma_3$. Bob's measurement choices are $\hat{B}_0^{(3)}=I$ and $\hat{B}_1^{(3)}=\sigma_1$, and the corresponding unitary operations are $\hat{U}_y^{(3)}=I$. Charlie's measurement choices are $\hat{C}_0^{(3)}=\frac{\sqrt{2}}{2}\sigma_1+\frac{\sqrt{2}}{2}\sigma_3$ and $\hat{C}_1^{(3)}=\frac{\sqrt{2}}{2}\sigma_1-\frac{\sqrt{2}}{2}\sigma_3$. We obtain $S_{L_1}^{(3)}=\frac{1}{\sqrt{2}}\left|\cos\delta\right|,S_{L_2}^{(3)}=\frac{1}{2}\left|2\cos\delta+\sin\delta\right|$. Within range of $S_{L_1}^{(3)}\in\left[\sqrt{\frac{2}{5}},\frac{1}{\sqrt{2}}\right]$, the trade-off is $S_{L_2}^{(3)}=\sqrt{2}S_{L_1}^{(3)}+\frac{1}{2}\sqrt{1-2\left(S_{L_1}^{(3)}\right)^2}$. $S_{L_2}^{(3)}$ reaches its maximum value $\frac{\sqrt{5}}{2}$ at $S_{L_1}^{(3)}=\sqrt{\frac{2}{5}}$.
	
\begin{figure}[htbp]
		\vspace{-2mm} 
		\begin{center}
			\includegraphics[width=\columnwidth]{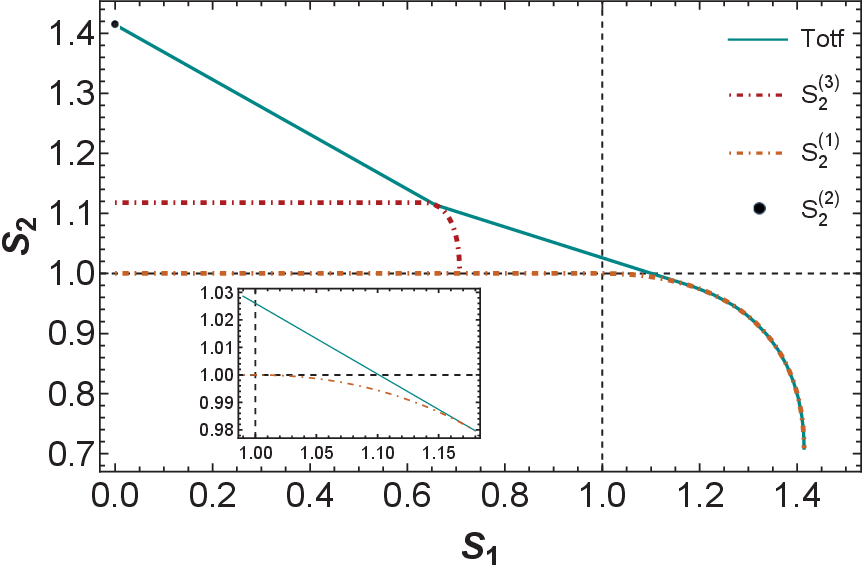} 
			
			\caption{Illustration of quantum steering sharing by two linear steering inequalities. Plot all trade-offs: the optimal trade-off Totf (dark green line), $S_2^{(3)}$ (red dot-dashed line), $S_2^{(1)}$ (orange dot-dashed line), and $S_2^{(2)}$ (black dot). The black dashed line represents the classical bound 1 of the linear steering inequality. The inserted figure at the left bottom corner highlights the region where two linear inequalities are simultaneously violated, which is a closed region enclosed by the dark green solid curve and two classical boundaries (two black dashed lines).} \label{figure2}
		\end{center}
		\vspace{-5mm}  
	\end{figure}

	From the above results, we clearly observe that the double violation of the linear steering inequalities is not possible in cases 1, 2, and 3, individually.  However, the double violation is attainable by stochastically combining these three cases using the probability distribution $\{p\left(\lambda\right)\}_{\lambda=1}^3$. First, we combine Case 1 and Case 3 with a straight line. When this line is tangent to both Case 1 and Case 3, we obtain the optimal combination of Case 1 and Case 3. Then, we draw a tangent line to Case 3 that passes through the point of Case 2 to obtain the optimal combination of Case 2 and Case 3. In this way, we have calculated that the tangent point obtained by combining Case 2 and Case 3 is to the right of the tangent point to the left of the tangent line
	obtained by combining Case 1 and Case 3. So the optimal trade-off which is defined to be the maximum possible value of $S_2$ for a given value of $S_1$, consists of three boundary regions: a mixture of Case 2 and Case 3, a mixture of Case 1 and Case 3, and Case 1 individually. The optimal trade-off is numerically found as 
	\begin{align}
		S_2=\begin{cases}
			\left(\sqrt{2}-\sqrt{\frac{7}{2}}\right)S_1+\sqrt{2}&\text{ if }0\le S_1\le0.656,    \\
			-0.257S_1+1.283&\text{ if }0.656\le S_1\le1.180,\\
			\frac{1}{2}\left(S_1+\sqrt{2-\left(S_1\right)^2}\right)& \text{ if } 1.180\le S_1\le\sqrt{2}.   
		\end{cases}\label{d}
	\end{align} 
	The optimal trade-off (\ref{d}) is also plotted in Fig. \ref{figure2}. The area bounded by the dark green line and two black dashed lines in Fig. \ref{figure2} represents the region in which the two linear steering inequalities are violated at the same time. The maximum double violation of the linear steering inequalities is $S_1=S_2\approx1.021$. Subsequently, we can obtain the optimal measurement settings that achieve the maximum double violation via the numerical search method, which maximizes the value of $\min\{S_1, S_2\}$. Specifically, when stochastically combining Case 1 with Case 3 at $\theta=-\frac{\pi}{15},\delta=\frac{\pi}{8}$, the maximum double violation is $S_1=S_2\approx1.021$. The probability of Case 1 is $p\left(1\right)\approx 0.690$ and hence $p\left(2\right)=0$ and $p\left(3\right)\approx 0.310$. When stochastically combining Case 1 with Case 2 at $\theta=-\frac{\pi}{12},\delta=0$, the maximum double violation is $S_1=S_2\approx1.035$. The probability of Case 1 is $p\left(1\right)\approx 0.845$ and so $p\left(2\right)\approx 0.155$ and $p\left(3\right)=0$. Moreover,
	if given a partially entangled state of the form $\left|\phi\right\rangle=\cos\alpha\left|00\right\rangle+\sin\alpha\left|11\right\rangle$, the maximum double violation is $S_1=S_2\approx1.021$ at $\alpha=\frac{23\pi}{96}$ by stochastically combining Case 1 with Case 3. And the maximum double violation is $S_1=S_2\approx1.043$ at $\alpha=\frac{7\pi}{36}$ by stochastically combining Case 1 with Case 2. Therefore, it is clearly shown that quantum steering can be recycled using only standard projective measurements. More details of the analysis are given in \hyperref[A]{Appendix A}. However, the double violations found in this situation are quite weak, compared to the violation $S_1=S_2=1.131$ (see \hyperref[C]{Appendix C} for details). This is because two-settings linear steering inequality has low robustness \cite{Qu.Rui.PhysRevLett.128.240402_2022,Zeng.Qiang.PhysRevLett.120.030401_2018}. As a result, the projective measurement protocol imposes a non-negligible impact on the violation of two-settings linear steering inequality. So the violation of the linear steering inequalities decreases rapidly.
	\subsection{The case of the CHSH inequality and one linear steering inequality} \label{3}
	We next discuss an interesting quantum correlation sharing and name it hybrid quantum correlation sharing, which can be characterized by the double violation of the CHSH inequality and one linear steering inequality. Specifically, the hybrid quantum correlation can be tested by the CHSH inequality (between Alice and Bob/Charlie) and the linear steering inequality (between Alice and Charlie/Bob). 
	
	We first discuss the CHSH inequality for Alice and Bob and the linear steering inequality for Alice and Charlie, expressing the CHSH inequality and the linear steering inequality as $S_{C_1}=\frac{1}{2}(\left\langle A_0\otimes B_0\right\rangle+\left\langle A_0\otimes B_1\right\rangle+ \left\langle A_1\otimes B_0\right\rangle-\left\langle A_1\otimes B_1\right\rangle)\le1, 
	S_{L_2}=\frac{1}{\sqrt{2}}\left|\left\langle A_0\otimes C_0\right\rangle+\left\langle A_1\otimes C_1\right\rangle\right|\le1$. Here, $S_{C_1}$ and $S_{L_2}$ denote the CHSH and the linear steering parameter, respectively. Since the classical bound of the CHSH inequality is 2 and the classical bound of the linear steering inequality is 1,  to better explore the simultaneous violation of the CHSH inequality and the linear steering inequality, the CHSH inequality is weighted such that its classical bound is also 1. For the maximally entangled state $\rho$, we explain the optimal measurement settings for three cases of measurements in the following.
	
	Case 1. When $\lambda=1$, both measurements are basis projections for Bob. Alice's measurement choices are $\hat{A}_0^{(1)}=\cos\mu\sigma_1+\sin\mu\sigma_3$ and $\hat{A}_1^{(1)}=\cos\mu\sigma_1-\sin\mu\sigma_3$. Bob's measurement choices are $\hat{B}_0^{(1)}=\sigma_1$ and $\hat{B}_1^{(1)}=\cos(2\mu)\sigma_1+\sin(2\mu)\sigma_3$, and the corresponding unitary operations are $\hat{U}_0^{(1)}=I, \hat{U}_1^{(1)}=e^{-i\mu\sigma_2}$. Charlie's measurement choices are $\hat{C}_0^{(1)}=\frac{\sqrt{2}}{2}\sigma_1-\frac{\sqrt{2}}{2}\sigma_3, \hat{C}_1^{(1)}=\frac{\sqrt{2}}{2}\sigma_1+\frac{\sqrt{2}}{2}\sigma_3$. We obtain $S_{C_1}^{(1)}=\frac{1}{2}(3\cos\mu-\cos(3\mu))$,  $S_{L_2}^{(1)}=\left|\cos\mu\right|^3$. The trade-off is $S_{L_2}^{(1)}=\frac{\left[2^{1/3}+(-S_{C_1}^{(1)}+\sqrt{-2+(S_{C_1}^{(1)})^2})^{2/3}\right]^3}{4(-S_{C_1}^{(1)}+\sqrt{-2+(S_{C_1}^{(1)})^2})}$ in the range of $S_{C_1}^{(1)}\in\left[1,\sqrt{2}\right]$ and we easily obtain the classic trade-off $S_{L_2}^{(1)}=1$ in the range of $S_{C_1}^{(1)}\in\left[0,1\right]$.
	
	Case 2. When $\lambda=2$, both measurements are identity measurements for Bob. Alice's measurement choices are $\hat{A}_0^{(2)}=\frac{\sqrt{2}}{2}\sigma_1+\frac{\sqrt{2}}{2}\sigma_3$ and $\hat{A}_1^{(2)}=\frac{\sqrt{2}}{2}\sigma_1-\frac{\sqrt{2}}{2}\sigma_3$. Bob's measurement choices are $\hat{B}_0^{(2)}=\hat{B}_1^{(2)}=I$, and the corresponding unitary operations are $\hat{U}_y^{(2)}=I$. Charlie's measurement choices are $\hat{C}_0^{(2)}=\frac{\sqrt{2}}{2}\sigma_1+\frac{\sqrt{2}}{2}\sigma_3$ and $\hat{C}_1^{(2)}=\frac{\sqrt{2}}{2}\sigma_1-\frac{\sqrt{2}}{2}\sigma_3$. We obtain $S_{C_1}^{(2)}=0$, $S_{L_2}^{(2)}=\sqrt{2}$, where $S_{L2}^{(2)}$ reaches the quantum bound of the linear steering inequality.
	
	Case 3. When $\lambda=3$, one measurement is a basis projection and the other is an identity measurement for Bob. Alice's measurement choices are $\hat{A}_0^{(3)}=\cos\nu\sigma_1+\sin\nu\sigma_3$,  $\hat{A}_1^{(3)}=\cos\nu\sigma_1-\sin\nu\sigma_3$. Bob's measurement choices are $\hat{B}_0^{(3)}=I, \hat{B}_1^{(3)}=\sigma_3$, and the corresponding unitary operations are $\hat{U}_y^{(3)}=I$. Charlie's measurement choices are $\hat{C}_0^{(3)}=\frac{\sqrt{2}}{2}\sigma_1+\frac{\sqrt{2}}{2}\sigma_3,  \hat{C}_1^{(3)}=\frac{\sqrt{2}}{2}\sigma_1-\frac{\sqrt{2}}{2}\sigma_3$. We obtain $S_{C_1}^{(3)}=\sin\nu, S_{L_2}^{(3)}=\frac{1}{2}\left|\cos\nu+2\sin\nu\right|$. Within range of $S_{L_1}^{(3)}\in\left[\frac{2}{\sqrt{5}},1\right]$, the trade-off is $S_{L_2}^{(3)}=\frac{\sqrt{1-(S_{C_1}^{(3)})^2}}{2}+S_{C_1}^{(3)}$. $S_{L_2}^{(3)}$ reaches its maximum value $\frac{\sqrt{5}}{2}$ at $S_{C_1}^{(3)}=\frac{2}{\sqrt{5}}$.
	
\begin{figure}[htbp]
		\vspace{-2mm} 
		\begin{center}
			\includegraphics[width=\columnwidth]{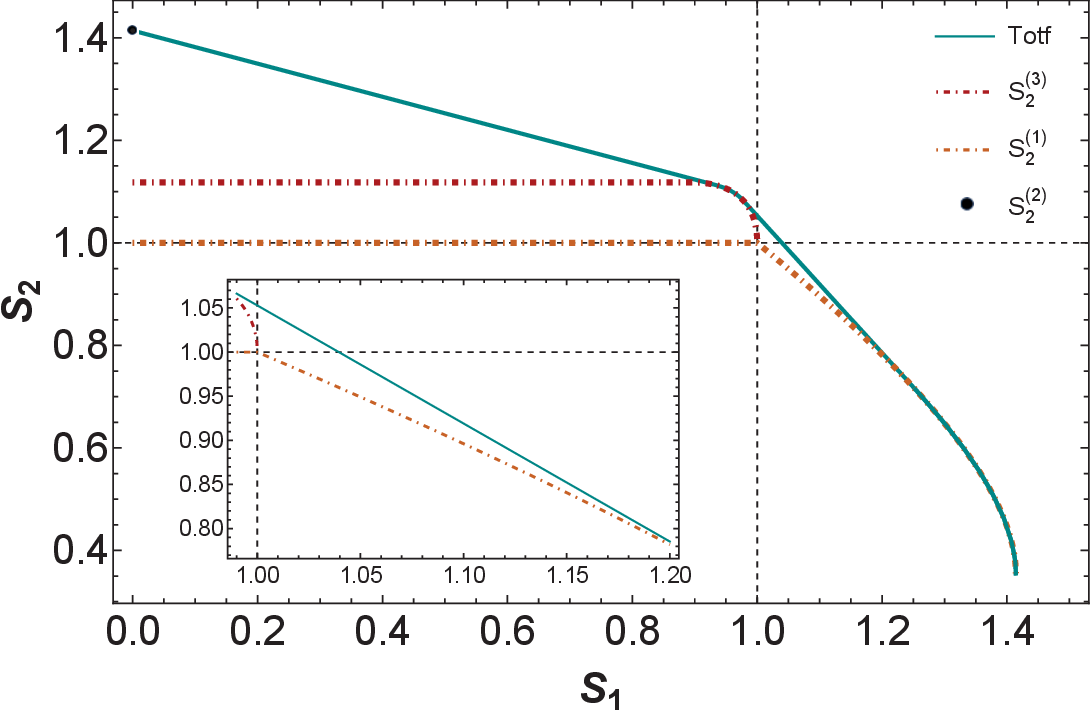} 
			
			\caption{Characterizing hybrid quantum steering sharing from the CHSH inequality and the linear steering inequality. Plot all trade-offs: the optimal trade-off Totf (dark green line), $S_2^{(3)}$ (red dot-dashed line), $S_2^{(1)}$ (orange dot-dashed line), and $S_2^{(2)}$ (black dot). The black dashed line represents the classical bound 1 of the linear steering inequality. The inserted figure at the left bottom corner indicates the double violation region of the CHSH inequality (between Alice and Bob) and the linear steering inequality (between Alice and Charlie).} \label{figure 3}
		\end{center}
		\vspace{-5mm}  
	\end{figure}

	Similarly, we introduce randomness described by $\{p\left(\lambda\right)\}_{\lambda=1}^3$ to stochastically combine these three cases to obtain the double violation of the CHSH inequality (between Alice and Bob) and the linear steering inequality (between Alice and Charlie). It is easy to calculate that the tangent point obtained by combining Case 2 and Case 3 is to the left of the tangent point to the left of the tangent line obtained by combining Case 1 and Case 3. Therefore, the optimal trade-off for the maximally entangled state $\rho$, consists of four
	boundary regions: a mixture of Case 2 and Case 3, Case 3 individually, a mixture of Case 1 and Case 3, and Case 1 individually. The optimal trade-off is numerically calculated as,
	\begin{align}
		S_2=\begin{cases}
			\left(1-\frac{\sqrt{7}}{2}\right)S_1+\sqrt{2}&\text{ if }0\le S_1\le\sqrt{\frac{7}{8}}    \\
			\frac{\sqrt{1-(S_1)^2}}{2}+S_1&\text{ if }\sqrt{\frac{7}{8}}\le S_1\le 0.978\\
			-1.337S_1+2.390&\text{ if }0.978\le S_1\le1.254\\
			\frac{\left[2^{1/3}+(-S_1+\sqrt{-2+(S_1)^2})^{2/3}\right]^3}{4(-S_1+\sqrt{-2+(S_1)^2})}& \text{ if } 1.254\le S_1\le\sqrt{2}   
		\end{cases}\label{h}
	\end{align}
	As illustrated in Fig. \ref{figure 3}, the simultaneous violation region of the CHSH inequality (between Alice and Bob) and the linear steering inequality (between Alice and Charlie) is enclosed by the dark green line and two black dashed lines. The maximum double violation of the CHSH inequality (between Alice and Bob) and the linear steering inequality (between Alice and Charlie) is approximately 1.023. Then using the numerical search method, we can also obtain the optimal measurement settings that maximize the double violation in different combinations. When stochastically combining Case 1 with Case 3 at $\mu=\frac{7\pi}{47},\nu=\frac{27\pi}{62}$, the maximum double violation is $S_1=S_2\approx1.022$, where the desired probability distribution is $p\left(1\right)\approx 0.156$, $p\left(2\right)=0$ and $p\left(3\right)\approx 0.844$. When stochastically combining Case 1 with Case 2, the maximum value of $S_1$ and $ S_2$ is always 1. Without loss of generality, we also consider a partially entangled state given as $\left|\phi\right\rangle$, where $\alpha\in[0,\frac{\pi}{4}]$. The common maximum value of $S_1$ and $S_2$ is approximately 1.030 at $\alpha=\frac{2\pi}{9}$ when stochastically combining Case 1 with Case 3. We obtain $S_1=S_2=1$ at $\alpha=\frac{\pi}{4}$ when stochastically combining Case 1 with Case 2 (see \hyperref[A]{Appendix A} for details). It is clearly found that the hybrid quantum correlation sharing can be observed when stochastically combining Case 1 and Case 3. However, in this study, the maximum value of the double violation is still weak, compared to the violation $S_1=S_2=1.131$ (see \hyperref[C]{Appendix C} for details).
	
	%However, the maximum double violation of the CHSH inequality (between Alice and Bob) and the linear steering inequality (between Alice and Charlie) under the projective measurement strategy cannot reach the violation $S_1=S_2\approx1.131$ under the weak measurement (\hyperref[D]{Appendix D}). 
	
	On the other hand, the CHSH inequality can be used for Alice and Charlie and the linear steering inequality can be used for Alice and Bob, with $S_{L_1}=\frac{1}{\sqrt{2}}\left|\left\langle A_0\otimes B_0\right\rangle+\left\langle A_1\otimes B_1\right\rangle\right|\le1, S_{C_2}=\frac{1}{2}(\left\langle A_0\otimes C_0\right\rangle+\left\langle A_0\otimes C_1\right\rangle+\left\langle A_1\otimes C_0\right\rangle-\left\langle A_1\otimes C_1\right\rangle)\le1$. Here, $S_{L_1}$ and $S_{C_2}$ show the linear steering and the CHSH parameter, respectively. There exist three trade-offs for the maximally entangled state $\rho$ under the projective measurement strategies. Unfortunately, we find the common maximum value of $S_1$ and $S_2$ is always less than 1, regardless of stochastically combining Case 1 and Case 2 or Case 1 and Case 3. Similar results can be observed when considering partially entangled state $\left|\phi\right\rangle$, and it is not possible to find the double violation of the CHSH inequality (between Alice and Charlie) and the linear steering inequality (between Alice and Bob) for any choice of $\alpha$. The reason for such a result may be that the performance of quantum steering verified by linear steering inequality is so weak.
	
	\section{Bob and Charlie to steer Alice's state}
	In particular, quantum steering is directional, with one party steering the other, but not vice versa \cite{Bowles.PhysRevLett.112.200402_2014}. Therefore, we study the ability of Bob and Charlie to simultaneously steer Alice's state. %\textcolor{red}{To distinguish between Alice's two measurement settings, the measurement directions of Alice are set to be vertical to one another.} 
	Meanwhile, the two observables of Bob and Charlie are defined in a general form.
	\subsection{The case of two linear steering inequalities} \label{4}
	
	Considering the maximally entangled state $\rho$, we list out the optimal measurement settings for three cases of projective measurements below.  
	
	Case 1. When $\lambda=1$, both measurements are basis projections for Bob. Alice's measurement choices are $\hat{A}_0^{(1)}=\frac{\sqrt{2}}{2}\sigma_1-\frac{\sqrt{2}}{2}\sigma_3$ and $\hat{A}_1^{(1)}=\frac{\sqrt{2}}{2}\sigma_1+\frac{\sqrt{2}}{2}\sigma_3$. Bob's measurement choices are $\hat{B}_0^{(1)}=\cos\chi\sigma_1+\sin\chi\sigma_3$ and $\hat{B}_1^{(1)}=\cos\chi\sigma_1-\sin\chi\sigma_3$, and the corresponding unitary operations are $\hat{U}_0^{(1)}=I, \hat{U}_1^{(1)}=e^{i\chi\sigma_2}$. Charlie's measurement choices are $\hat{C}_0^{(1)}=\hat{C}_1^{(1)}=\cos\chi\sigma_1+\sin\chi\sigma_3$.  Under these measurement settings, we obtain $S_{L_1}^{(1)}=\left|\cos\chi-\sin\chi\right|$, $S_{L_2}^{(1)}=\left|\cos\chi\right|$. The trade-off is $S_{L_2}^{(1)}=\frac{1}{2}\left(S_{L_1}^{(1)}+\sqrt{2-\left(S_{L_1}^{(1)}\right)^2}\right)$ in the range of $S_{L_1}^{(1)}\in\left[1,\sqrt{2}\right]$ and we easily obtain the classic trade-off $S_{L_2}^{(1)}=1$ in the range of $S_{L_1}^{(1)}\in\left[0,1\right]$.
	
	Case 2. When $\lambda=2$, both measurements are identity measurements for Bob. Alice's measurement choices are $\hat{A}_0^{(2)}=\frac{\sqrt{2}}{2}\sigma_1-\frac{\sqrt{2}}{2}\sigma_3$ and $\hat{A}_1^{(2)}=\frac{\sqrt{2}}{2}\sigma_1+\frac{\sqrt{2}}{2}\sigma_3$. Bob's measurement choices are $\hat{B}_0^{(2)}=\hat{B}_1^{(2)}=I$, and the corresponding unitary operations are $\hat{U}_y^{(2)}=I$. Charlie's measurement choices are $\hat{C}_0^{(2)}=\frac{\sqrt{2}}{2}\sigma_1-\frac{\sqrt{2}}{2}\sigma_3$ and $\hat{C}_1^{(2)}=\frac{\sqrt{2}}{2}\sigma_1+\frac{\sqrt{2}}{2}\sigma_3$. We obtain $S_{L1}^{(2)}=0$, $S_{L2}^{(2)}=\sqrt{2}$, where $S_{L2}^{(2)}$ reaches the quantum bound of the linear steering inequality.
	
	Case 3a. When $\lambda=3$, one measurement is a basis projection and the other is an identity measurement for Bob. Alice's measurement choices are $\hat{A}_0^{(3)}=\frac{\sqrt{2}}{2}\sigma_1-\frac{\sqrt{2}}{2}\sigma_3$ and $\hat{A}_1^{(3)}=\frac{\sqrt{2}}{2}\sigma_1+\frac{\sqrt{2}}{2}\sigma_3$. Bob's measurement choices are $\hat{B}_0^{(3)}=I$ and $ \hat{B}_1^{(3)}=\cos\omega\sigma_1+\sin\omega\sigma_3$, and the corresponding unitary operations are $\hat{U}_y^{(3)}=I$. Charlie's measurement choices are $\hat{C}_0^{(3)}=\frac{2}{\sqrt{5}}\sigma_1-\frac{1}{\sqrt{5}}\sigma_3$ and $\hat{C}_1^{(3)}=\frac{2}{\sqrt{5}}\sigma_1+\frac{1}{\sqrt{5}}\sigma_3$. We obtain $S_{L_1}^{(3)}=\frac{1}{2}\left|\sin\omega+\cos\omega\right|,S_{L_2}^{(3)}=\left|\frac{9+\cos(2\omega)}{4\sqrt{5}}\right|$. Within range of $S_{L_1}^{(3)}\in\left[\frac{1}{2},\frac{1}{\sqrt{2}}\right]$, the trade-off is $S_{L_2}^{(3)}=\frac{9+2S_{L_1}^{(3)}\sqrt{2-4(S_{L_1}^{(3)})^2}}{4\sqrt{5}}$. $S_{L_2}^{(3)}$ reaches its maximum value $\frac{\sqrt{5}}{2}$ at $S_{L_1}^{(3)}=\frac{1}{2}$. 
	
	Case 3b. For $\lambda=3$, there are other optimal measurement settings available. Alice's measurement choices are $\hat{A}_0^{(3)}=\cos\varpi\sigma_1+\sin\varpi\sigma_3$ and $\hat{A}_1^{(3)}=-\sin\varpi\sigma_1+\cos\varpi\sigma_3$. Bob's measurement choices are $\hat{B}_0^{(3)}=I$ and $\hat{B}_1^{(3)}=\sigma_1$, and the corresponding unitary operations are $\hat{U}_y^{(3)}=I$. Charlie's measurement choices are $\hat{C}_0^{(3)}=\frac{2}{\sqrt{5}}\sigma_1-\frac{1}{\sqrt{5}}\sigma_3$ and $\hat{C}_1^{(3)}=\frac{2}{\sqrt{5}}\sigma_1+\frac{1}{\sqrt{5}}\sigma_3$. We obtain $S_{L_1}^{(3)}=\frac{1}{\sqrt{2}}\left|\sin\varpi\right|,S_{L_2}^{(3)}=\frac{\sqrt{10}}{4}\left|\cos\varpi-\sin\varpi\right|$. Within range of $S_{L_1}^{(3)}\in\left[\frac{1}{2},\frac{1}{\sqrt{2}}\right]$, the trade-off is $S_{L_2}^{(3)}=\frac{\sqrt{5}}{4}(2S_{L_1}^{(3)}+\sqrt{2-4(S_{L_1}^{(3)})^2})$. $S_{L_2}^{(3)}$ reaches its maximum value $\frac{\sqrt{5}}{2}$ at $S_{L_1}^{(3)}=\frac{1}{2}$. 
	
\begin{figure}[htbp]
		\vspace{-2mm} 
		\begin{center}
			\includegraphics[width=\columnwidth]{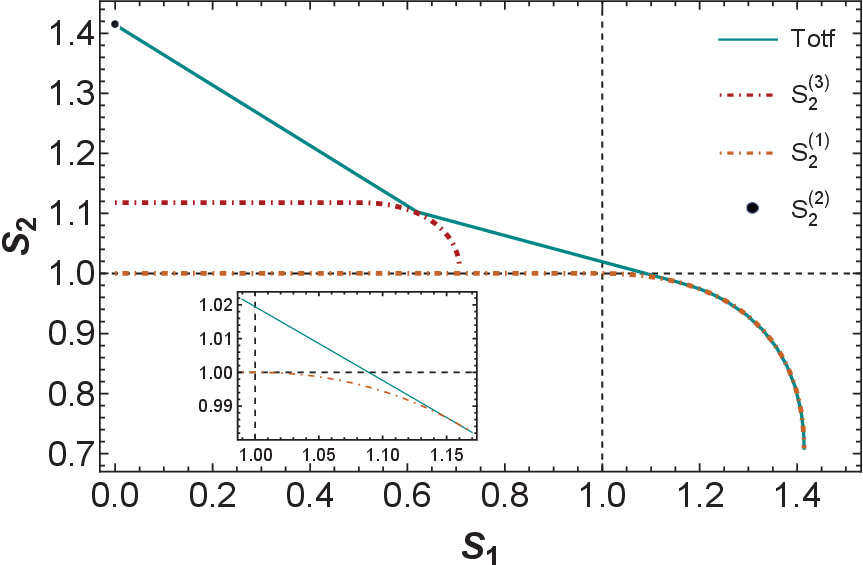} 
			
			\caption{Revealing quantum steering sharing with Case 3a by resorting to two linear steering inequalities. The dark green line represents the optimal trade-off, which is obtained through a rigorous analysis for the maximally entangled state based on standard projective measurements. The inserted figure at the left bottom corner illustrates the double violation region of the linear steering inequalities.} \label {figure 4}
		\end{center}
		\vspace{-5mm}  
	\end{figure}

	Obviously, we have two analytical results when $\lambda=3$. Let's first discuss the former result shown in Case 3a. Since the tangent
	point obtained by combining Case 2 and Case 3 is greater than the tangent point to the left of the tangent line obtained by combining Case 1 and Case 3, the optimal trade-off consists of three boundary regions: a mixture of Case 2 and Case 3, a mixture of Case 1 and Case 3, and Case 1 individually. The optimal trade-off can be obtained as
	\begin{align}
		S_2=\begin{cases}
			-0.503S_1+\sqrt{2}&\text{ if }0\le S_1\le0.619    \\
			-0.218S_1+1.238&\text{ if }0.619\le S_1\le1.161\\
			\frac{1}{2}\left(S_1+\sqrt{2-\left(S_1\right)^2}\right)& \text{ if } 1.161\le S_1\le\sqrt{2}   
		\end{cases}\label{j}
	\end{align}
	It is shown in Fig. \ref{figure 4} that there exists the double violation of the linear steering inequalities, which is indicated by the region bounded by the dark green line and two black dashed lines, and the maximum violation of the linear steering inequalities is approximately 1.016. Using the numerical search method, we obtain the optimal measurement settings that maximize the double violation of the linear steering inequalities. The maximum double violation of the linear steering inequalities is approximately 1.016 when stochastically combining Case 1 with Case 3 at $\chi=-\frac{\pi}{18},\omega=\frac{\pi}{16}$, with the probability of Case 1 as $p\left(1\right)\approx 0.750$. While for the stochastical combination of Case 1 and Case 2 at $\chi=-\frac{\pi}{12}, \omega=0$, the maximum value of $S_1$ and $S_2$ is approximately 1.035, where the probability of Case 1 is $p\left(1\right)\approx 0.845$. For the partially entangled state $\left|\phi\right\rangle$, the maximum double violation of the linear steering inequalities is $S_1=S_2\approx1.017$ at $\alpha=\frac{10\pi}{37}$ by randomly mixing Cases 1 and 3. And the maximum double violation of the linear steering inequalities is $S_1=S_2\approx1.043$ at $\alpha=\frac{7\pi}{36}$ by randomly mixing Cases 1 and 2 (see \hyperref[B]{Appendix B} for details).
	
\begin{figure}[htbp]
		\vspace{-2mm} 
		\begin{center}
			\includegraphics[width=\columnwidth]{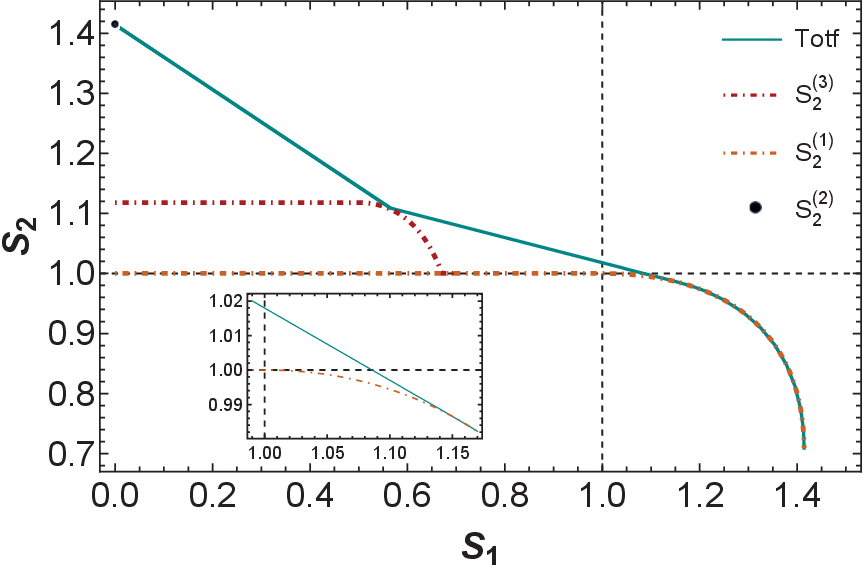}
			
			\caption{Demonstration of quantum steering sharing with Case 3b by two linear steering inequalities. The dark green line represents the variation of the optimal trade-off values that are obtained for the maximally entangled state using standard projective measurements. The double violation region is specified in the inserted figure at the left bottom corner.} 
			\label{figure 5}
		\end{center}
		\vspace{-5mm}  
	\end{figure}

	Moreover, we investigate the double violation of the linear steering inequalities with the measurement settings given in Case 3b. The tangent point obtained by combining Case 2 and Case 3 is also greater than the tangent point to the left of the tangent line obtained by combining Case 1 and Case 3. Therefore the optimal trade-off consists of three boundary regions: a mixture of Case 2 and Case 3, a mixture of Case 1 and Case 3, and Case 1 individually. We give the optimal trade-off as 
	\begin{align}
		S_2=\begin{cases}
			\frac{\sqrt{5}}{2}(1-\sqrt{\frac{11}{5}})S_1+\sqrt{2}&\text{ if }0\le S_1\le0.565\\
			-0.209S_1+1.227&\text{ if }0.565\le S_1\le1.156\\
			\frac{1}{2}\left(S_1+\sqrt{2-\left(S_1\right)^2}\right)& \text{ if } 1.156\le S_1\le\sqrt{2}   
		\end{cases}\label{k}
	\end{align}
	In Fig. \ref{figure 5}, we demonstrate the double violation by plotting the trade-off values. The common maximum value of $S_1$ and $S_2$ is approximately 1.016 from calculations. We then also obtain the optimal measurement settings by means of numerical searching. The maximum double violation is $S_1=S_2\approx1.015$ when integrating Case 1 and Case 3 in a random manner with $p\left(1\right)\approx 0.767$ at $\chi=-\frac{\pi}{18},\varpi=-\frac{5\pi}{18}$. And the maximum double violation is roughly 1.035 when merging Case 1 and Case 2 randomly, with $p\left(1\right)\approx 0.845$ at $\chi=-\frac{\pi}{12},\varpi=0$. Given the partially entangled state $\left|\phi\right\rangle$, the maximum double violation is $S_1=S_2\approx1.016$ at $\alpha=\frac{10\pi}{37}$ based on the random combination of Case 1 and Case 3. And the maximum value of $S_1$ and $S_2$ is approximately 1.043 at $\alpha=\frac{7\pi}{36}$ with Case 1 and Case 2 integrated at random (see \hyperref[B]{Appendix B} for details). 
	Similarly, the violation obtained under projective measurements is smaller, compared to the violation $S_1=S_2\approx1.131$ obtained under the weak measurement (see \hyperref[C]{Appendix C} for details). 
	
	\subsection{The case of the CHSH inequality and one linear steering inequality} \label{5}
	The hybrid quantum correlation sharing controlled by Bob and Charlie cannot be achieved in our protocol. Firstly, we investigate the CHSH inequality between Alice and Bob, and the linear steering inequality between Alice and Charlie. We find three trade-offs for the maximally entangled state with the projective measurement strategies. Regardless of stochastically combining Case 1 and Case 2 or Case 1 and Case 3, we cannot observe any violation with $S_1=S_2<1$. For the partially entangled state $\left|\phi\right\rangle$, where $\alpha\in[0,\frac{\pi}{4}]$, we also obtain $S_1=S_2<1$. Secondly, the CHSH inequality between Alice and Charlie and the linear steering inequality between Alice and Bob are explored. Neither the maximally entangled state $\rho$ nor the partially entangled state $\left|\phi\right\rangle$ can demonstrate the hybrid quantum correlation sharing.
	
	\section{Conclusion}
	In this work, for the maximally entangled states $\rho$, we realize Alice can simultaneously steer the states of Bob and Charlie and conversely, Bob and Charlie simultaneously steer Alice's state through the standard projective measurements. We extend this analysis to partially entangled states $\left|\phi\right\rangle$, where quantum steering sharing can still be observed. The maximum double violation of the linear steering inequality obtained from partially entangled states can be greater in some cases than that obtained from maximally
	entangled states when randomly mixing Case 1 and Case 2. Moreover, the hybrid quantum correlation sharing is discussed by the double violation of the CHSH inequality (between Alice and Bob) and one linear steering inequality (between Alice and Charlie). we have numerically demonstrated that the hybrid quantum correlation sharing can be realized in our protocol. However, the whole work indicates that the results obtained using the projective measurement strategies consistently fall short of the violation achieved under weak measurement. 
	
	\emph{Discussion-}
	We have achieved the quantum steering sharing of Bob and Charlie with Alice through the standard projective measurements, but there are still some unexplored research areas. Firstly, is it possible to potentially reduce the weakening effect of the projective measurement strategies on the linear steering inequality by increasing the number of observer's measurement settings? Moreover, since it has been proven in the unilateral sequential scenario that Alice can steer the states of multiple Bobs \cite{Yao.Dan.PhysRevA.103.052207_2021} and multiple Bobs can steer Alice's state \cite{Shenoy.H.PhysRevA.99.022317_2019}, it remains open questions whether Alice can steer the states of multiple Bobs and how many Bobs can steer Alice's state through the standard projective measurement strategies. Additionally, in the bipartite scenario, Zhu et al. \cite{Zhu.Jie.PhysRevA.105.032211_2022} have already achieved quantum steering sharing. 
	Is it possible to achieve bilateral quantum steering sharing under the standard projective measurement strategies? These questions can give us ideas to better study quantum steering.
	
	\section{Acknowledgment}
	
	C.R. was supported by the National Natural Science Foundation of China (Grant No. 12075245, 12247105), the Natural Science Foundation of Hunan Province (2021JJ10033), Xiaoxiang Scholars Programme of Hunan Normal University, the Foundation Xiangjiang-Laboratory (XJ2302001) and Hunan provincial major sci-tech program No. (2023zk1010).  C.W. is supported by the National Research Foundation, Singapore and A*STAR under its Quantum Engineering Programme (NRF2021-QEP2-02-P03).
	
	\bibliographystyle{apsrev4-1}
	\bibliography{ref}
	
	\appendix 
	\section{Alice steers Bob's and Charlie's states in partially entangled states} \label{A}
	We mainly discuss the ability of Alice to steer the states of Bob and Charlie under the partially entangled state $\left|\phi\right\rangle$. This state is defined as $\left|\phi\right\rangle=\cos\alpha\left|00\right\rangle+\sin\alpha\left|11\right\rangle$.  
	\subsection{The case of two linear steering inequalities}
	The optimal measurement settings are as follows: 
	
	Case 1. When $\lambda=1$, both measurements are basis projections for Bob. Alice's measurement choices are $\hat{A}_0^{(1)}=\cos\kappa\sigma_1+\sin\kappa\sigma_3$ and $\hat{A}_1^{(1)}=\cos\kappa\sigma_1-\sin\kappa\sigma_3$. Bob's measurement choices are $\hat{B}_0^{(1)}=\frac{\sqrt{2}}{2}\sigma_1-\frac{\sqrt{2}}{2}\sigma_3$ and $\hat{B}_1^{(1)}=\frac{\sqrt{2}}{2}\sigma_1+\frac{\sqrt{2}}{2}\sigma_3$, and the corresponding unitary operations are $\hat{U}_0^{(1)}=I, \hat{U}_1^{(1)}=e^{\frac{\pi}{4}i\sigma_2}$. Charlie's measurement choices are $\hat{C}_0^{(1)}=\frac{\sqrt{2}}{2}\sigma_1-\frac{\sqrt{2}}{2}\sigma_3$ and $\hat{C}_1^{(1)}=-\frac{\sqrt{2}}{2}\sigma_1+\frac{\sqrt{2}}{2}\sigma_3$. We obtain $S_{L_1}^{(1)}=\left|\cos\kappa\sin(2\alpha)-\sin\kappa\right|$, $S_{L_2}^{(1)}=\left|\sin\kappa\right|$.
	
	Case 2. When $\lambda=2$, both measurements are identity measurements for Bob. Alice's measurement choices are $\hat{A}_0^{(2)}=\frac{\sqrt{2}}{2}\sigma_1+\frac{\sqrt{2}}{2}\sigma_3$ and $\hat{A}_1^{(2)}=-\frac{\sqrt{2}}{2}\sigma_1+\frac{\sqrt{2}}{2}\sigma_3$. Bob's measurement choices are $\hat{B}_0^{(2)}=\hat{B}_1^{(2)}=I$, and the corresponding unitary operations are $\hat{U}_y^{(2)}=I$. Charlie's measurement choices are $\hat{C}_0^{(2)}=\frac{\sqrt{2}}{2}\sigma_1+\frac{\sqrt{2}}{2}\sigma_3$ and $\hat{C}_1^{(2)}=-\frac{\sqrt{2}}{2}\sigma_1+\frac{\sqrt{2}}{2}\sigma_3$. We obtain $S_{L_1}^{(2)}=\left|\cos(2\alpha)\right|$, $S_{L_2}^{(2)}=\left|\frac{1+\sin(2\alpha)}{\sqrt{2}}\right|$.
	
	Case 3. When $\lambda=3$, one measurement is a basis projection and the other is an identity measurement for Bob. Alice's measurement choices are $\hat{A}_0^{(3)}=\cos\tau\sigma_1+\sin\tau\sigma_3$ and $\hat{A}_1^{(3)}=\cos\tau\sigma_1-\sin\tau\sigma_3$. Bob's measurement choices are $\hat{B}_0^{(3)}=I, \hat{B}_1^{(3)}=\sigma_1$, and the corresponding unitary operations are $\hat{U}_y^{(3)}=I$. Charlie's measurement choices are $\hat{C}_0^{(3)}=\frac{\sqrt{2}}{2}\sigma_1+\frac{\sqrt{2}}{2}\sigma_3, \hat{C}_1^{(3)}=\frac{\sqrt{2}}{2}\sigma_1-\frac{\sqrt{2}}{2}\sigma_3$. We obtain $S_{L_1}^{(3)}=\left|\frac{\sin(2\alpha+\tau)}{\sqrt{2}}\right|,S_{L_2}^{(3)}=\frac{1}{2}\left|2\cos\tau\sin(2\alpha)+\sin\tau\right|$.  
	
	We then also obtain the optimal measurement settings employing the numerical search method. The maximum double violation is $S_1=S_2\approx1.021$ when integrating Case 1 and Case 3 in a random manner with $p(1)\approx0.719$ at $\kappa=-\frac{4\pi}{9},\tau=\frac{\pi}{8}$, $\alpha=\frac{23\pi}{96}$. And the maximum double violation is roughly 1.043 when merging Case 1 and Case 2 randomly, with $p(1)\approx0.795$ at $\kappa=-\frac{9\pi}{22},\tau=0$, $\alpha=\frac{7\pi}{36}$. 
	\subsection{The case of the CHSH inequality and one linear steering inequality}
	Similarly, we discuss the hybrid quantum correlation sharing using the CHSH inequality and the linear steering inequality under the partially entangled state $\left|\phi\right\rangle$ here, and the optimal measurement settings for three cases of projective measurements are shown below.
	
	Case 1. When $\lambda=1$, both measurements are basis projections for Bob. Alice's measurement choices are $\hat{A}_0^{(1)}=\cos\beta\sigma_1+\sin\beta\sigma_3$ and $\hat{A}_1^{(1)}=\cos\beta\sigma_1-\sin\beta\sigma_3$. Bob's measurement choices are $\hat{B}_0^{(1)}=\sigma_1$ and $\hat{B}_1^{(1)}=\cos(2\beta)\sigma_1+\sin(2\beta)\sigma_3$, and the corresponding unitary operations are $\hat{U}_0^{(1)}=I, \hat{U}_1^{(1)}=e^{-i\beta\sigma_2}$. Charlie's measurement choices are $\hat{C}_0^{(1)}=\frac{\sqrt{2}}{2}\sigma_1-\frac{\sqrt{2}}{2}\sigma_3, \hat{C}_1^{(1)}=\frac{\sqrt{2}}{2}\sigma_1+\frac{\sqrt{2}}{2}\sigma_3$. We obtain $S_{C_1}^{(1)}=\cos\beta(1+\sin(2\alpha)-\cos(2\beta))$, $S_{L_2}^{(1)}=\left|\sin(2\alpha)\cos^3\beta\right|$.
	
	Case 2. When $\lambda=2$, both measurements are identity measurements for Bob. Alice's measurement choices are $\hat{A}_0^{(2)}=\sigma_3$ and $\hat{A}_1^{(2)}=\sigma_1$. Bob's measurement choices are $\hat{B}_0^{(2)}=\hat{B}_1^{(2)}=I$, and the corresponding unitary operations are $\hat{U}_y^{(2)}=I$. Charlie's measurement choices are $\hat{C}_0^{(2)}=\sigma_3$ and $\hat{C}_1^{(2)}=\sigma_1$. We obtain $S_{C_1}^{(2)}=\cos(2\alpha)$, $S_{L_2}^{(2)}=\left|\frac{1+\sin(2\alpha)}{\sqrt{2}}\right|$.
	
	Case 3. When $\lambda=3$, one measurement is a basis projection and the other is an identity measurement for Bob. Alice's measurement choices are $\hat{A}_0^{(3)}=\cos\gamma\sigma_1+\sin\gamma\sigma_3$ and $\hat{A}_1^{(3)}=-\cos\gamma\sigma_1+\sin\gamma\sigma_3$. Bob's measurement choices are $\hat{B}_0^{(3)}=I, \hat{B}_1^{(3)}=\sigma_1$, and the corresponding unitary operations are $\hat{U}_y^{(3)}=I$. Charlie's measurement choices are $\hat{C}_0^{(3)}=\frac{\sqrt{2}}{2}\sigma_1+\frac{\sqrt{2}}{2}\sigma_3, \hat{C}_1^{(3)}=-\frac{\sqrt{2}}{2}\sigma_1+\frac{\sqrt{2}}{2}\sigma_3$. We obtain $S_{C_1}^{(3)}=\sin(2\alpha+\gamma), S_{L_2}^{(3)}=\frac{1}{2}\left|2\cos\gamma\sin(2\alpha)+\sin\gamma\right|$.  
	
	We then also obtain the optimal measurement settings by using the numerical search method. The maximum double violation of the CHSH inequality and the linear steering inequality is $S_1=S_2\approx1.030$ when stochastically combining Case 1 with Case 3 at $\beta=\frac{\pi}{6},\gamma=\frac{\pi}{11}$, $\alpha=\frac{2\pi}{9}$, with the probability of Case 1 as $p(1)\approx0.124$. Moreover, regardless of the angle $\{\beta,\gamma,\alpha\}$, the common maximum value of $S_1$ and $S_2$ remains constant at 1 by randomly mixing Case 1 and Case 2.  
	
	Besides, the hybrid quantum correlation sharing can also be tested by the double violation of the CHSH inequality (between Alice and Charlie) and the linear steering inequality (between Alice and Bob). There exist three trade-offs for the
	partially entangled state $\left|\phi\right\rangle$ under the standard projective measurements. Unfortunately, regardless of stochastically combining Case 1 and Case 2 or stochastically combining Case 1 and Case 3, we can only obtain $S_1=S_2<1$.
	\section{Bob and Charlie steer a single Alice in partially entangled states} \label{B}
	Similarly, we study the ability of Bob and Charlie to simultaneously steer Alice’s state under the partially entangled state $\left|\phi\right\rangle$. 
	
	\subsection{The case of two linear steering inequalities}
	The optimal measurement settings are as follows: 
	
	Case 1. When $\lambda=1$, both measurements are basis projections for Bob. Alice's measurement choices are $\hat{A}_0^{(1)}=\frac{\sqrt{2}}{2}\sigma_1-\frac{\sqrt{2}}{2}\sigma_3$ and $\hat{A}_1^{(1)}=\frac{\sqrt{2}}{2}\sigma_1+\frac{\sqrt{2}}{2}\sigma_3$. Bob's measurement choices are $\hat{B}_0^{(1)}=\cos\varepsilon\sigma_1+\sin\varepsilon\sigma_3$ and $\hat{B}_1^{(1)}=\cos\varepsilon\sigma_1-\sin\varepsilon\sigma_3$, and the corresponding unitary operations are $\hat{U}_0^{(1)}=I, \hat{U}_1^{(1)}=e^{(\frac{\pi}{2}+\varepsilon)i\sigma_2}$. Charlie's measurement choices are $\hat{C}_0^{(1)}=\cos\varepsilon\sigma_1+\sin\varepsilon\sigma_3$ and $\hat{C}_1^{(1)}=-\cos\varepsilon\sigma_1-\sin\varepsilon\sigma_3$. We obtain $S_{L_1}^{(1)}=\left|\cos\varepsilon\sin(2\alpha)-\sin\varepsilon\right|$, $S_{L_2}^{(1)}=\left|\sin\varepsilon\right|$.
	
	Case 2. When $\lambda=2$, both measurements are identity measurements for Bob. Alice's measurement choices are $\hat{A}_0^{(2)}=\frac{\sqrt{2}}{2}\sigma_1+\frac{\sqrt{2}}{2}\sigma_3$ and $\hat{A}_1^{(2)}=-\frac{\sqrt{2}}{2}\sigma_1+\frac{\sqrt{2}}{2}\sigma_3$. Bob's measurement choices are $\hat{B}_0^{(2)}=\hat{B}_1^{(2)}=I$, and the corresponding unitary operations are $\hat{U}_y^{(2)}=I$. Charlie's measurement choices are $\hat{C}_0^{(2)}=\frac{\sqrt{2}}{2}\sigma_1+\frac{\sqrt{2}}{2}\sigma_3$ and $\hat{C}_1^{(2)}=-\frac{\sqrt{2}}{2}\sigma_1+\frac{\sqrt{2}}{2}\sigma_3$. We obtain $S_{L_1}^{(2)}=\left|\cos(2\alpha)\right|$, $S_{L_2}^{(2)}=\left|\frac{1+\sin(2\alpha)}{\sqrt{2}}\right|$.
	
	Case 3a. When $\lambda=3$, one measurement is a basis projection and the other is an identity measurement for Bob. Alice's measurement choices are $\hat{A}_0^{(3)}=\frac{\sqrt{2}}{2}\sigma_1-\frac{\sqrt{2}}{2}\sigma_3$ and $\hat{A}_1^{(3)}=\frac{\sqrt{2}}{2}\sigma_1+\frac{\sqrt{2}}{2}\sigma_3$. Bob's measurement choices are $\hat{B}_0^{(3)}=I, \hat{B}_1^{(3)}=\cos\eta\sigma_1+\sin\eta\sigma_3$, and the corresponding unitary operations are $\hat{U}_y^{(3)}=I$. Charlie's measurement choices are $\hat{C}_0^{(3)}=\frac{2}{\sqrt{5}}\sigma_1-\frac{1}{\sqrt{5}}\sigma_3, \hat{C}_1^{(3)}=\frac{2}{\sqrt{5}}\sigma_1+\frac{1}{\sqrt{5}}\sigma_3$. We obtain $S_{L_1}^{(3)}=\frac{1}{2}\left|\sin(2\alpha)\cos\eta+\sin\eta-\cos(2\alpha)\right|, S_{L_2}^{(3)}=\left|\frac{3-\cos(2\eta)+6\sin(2\alpha)+\sin(2\alpha-2\eta)+\sin(2\alpha+2\eta)}{4\sqrt{5}}\right|$. 
	
	Case 3b. When $\lambda=3$, there are other optimal measurement settings available. Alice's measurement choices are $\hat{A}_0^{(3)}=\cos\zeta\sigma_1+\sin\zeta\sigma_3$ and $\hat{A}_1^{(3)}=-\sin\zeta\sigma_1+\cos\zeta\sigma_3$. Bob's measurement choices are $\hat{B}_0^{(3)}=I, \hat{B}_1^{(3)}=\sigma_1$, and the corresponding unitary operations are $\hat{U}_y^{(3)}=I$. Charlie's measurement choices are $\hat{C}_0^{(3)}=\frac{2}{\sqrt{5}}\sigma_1-\frac{1}{\sqrt{5}}\sigma_3, \hat{C}_1^{(3)}=\frac{2}{\sqrt{5}}\sigma_1+\frac{1}{\sqrt{5}}\sigma_3$. We obtain $S_{L_1}^{(3)}=\frac{1}{\sqrt{2}}\left|(\cos(2\alpha)-\sin(2\alpha))\sin\zeta\right|, S_{L_2}^{(3)}=\left|\frac{(1+4\sin(2\alpha))(\cos\zeta-\sin\zeta)}{2\sqrt{10}}\right|$. 
	
	Considering the former analytical result shown in Case 3a, the maximum double violation of the linear steering inequalities is $S_1=S_2\approx1.017$ when stochastically combining Case 1 with Case 3 at $\varepsilon=-\frac{15\pi}{34}, \eta=\frac{\pi}{15}$, $\alpha=\frac{10\pi}{37}$, with the probability of Case 1 as $p(1)\approx0.711$. While for the stochastical combination of Case 1 and Case 2 at $\varepsilon=-\frac{2\pi}{5},\eta=0$, $\alpha=\frac{7\pi}{36}$, the maximum double violation of the linear steering inequalities is $S_1=S_2\approx1.043$, where the probability of Case 1 is $p_1\approx0.780$. 
	
	Besides, considering the latter result given in Case 3b, the maximum double violation is $S_1=S_2\approx1.016$ when integrating Case 1 and Case 3 in a random manner with $p(1)\approx0.744$ at $\varepsilon=-\frac{4\pi}{9}, \zeta=-\frac{5\pi}{18}$, $\alpha=\frac{10\pi}{37}$. And the maximum double violation is roughly 1.043  when merging Case 1 and Case 2 randomly, with $p(1)\approx0.780$ at $\varepsilon=-\frac{2\pi}{5},\eta=0$, $\alpha=\frac{7\pi}{36}$.
	\subsection{The case of the CHSH inequality and one linear steering inequality}
	The results obtained from the partially entangled state $\left|\phi\right\rangle$ are similar to those obtained from the maximally entangled state $\rho$, and neither of them can achieve hybrid quantum correlation steering. When using the CHSH inequality (between Alice and Bob) and the linear steering inequality (between Alice and Charlie), there exist three trade-offs under the standard projective measurements. Regardless of randomly combining Case 1 and Case 2 or Case 1 and Case 3, we cannot observe any violation with $S_1=S_2<1$.  When using the CHSH inequality (between Alice and Charlie) and the linear steering inequality (between Alice and Bob), we also cannot observe the hybrid quantum correlation sharing.
	\section{Quantum steering sharing for the maximally entangled state through weak measurements} \label{C}
	Here, we adopt a weak measurement process similar to those previously used in \cite{Ren.Changliang.PhysRevA.100.052121_2019, Silva.PhysRevLett.114.250401_2015}. Considering the same scenario, Alice and Bob share the maximum entanglement state $\rho$. Alice performs strong measurements on one of the qubits. Bob then performs weak measurements on the other qubit and relays it to Charlie, who performs strong measurements. In this scenario, each observer independently performs two dichotomous measurements and the choice of measurement is unbiased. Alice's two observables are defined as $\hat{A}_x$, where $x\in\{0,1\}$ with  $\hat{A}_0=\cos\theta_1\sigma_1+\sin\theta_1\sigma_3$ and $  \hat{A}_1=\cos\theta_2\sigma_1+\sin\theta_2\sigma_3$. Bob's two observables are defined as $\hat{B}_y$, where $y\in\{0,1\}$ with $\hat{B}_0=\cos\theta_3\sigma_1+\sin\theta_3\sigma_3$ and $   \hat{B}_1=\cos\theta_4\sigma_1+\sin\theta_4\sigma_3$. And Charlie's two observables are defined as $\hat{C}_z$, where $z\in\{0,1\}$ with $\hat{C}_0=\cos\theta_5\sigma_1+\sin\theta_5\sigma_3$ and $ \hat{C}_1=\cos\theta_6\sigma_1+\sin\theta_6\sigma_3$. The corresponding outcomes are denoted as $\{a,b,c\}\in\{-1,1\}$. After the measurements of all observers are completed, we can obtain the joint probability distribution $P(a,b,c|x,y,z)$.  According to this joint probability distribution, we can obtain the marginal probability distribution $P(a,b|x,y)$ of the Alice-Bob pair, and the marginal probability distribution $P(a,c|x,z)$ of the Alice-Charlie pair. 
	
	To obtain the joint probability $P(a,b,c|x,y,z)$, we will introduce the measurement process. Alice performs a strong measurement $\hat{A}_x$ on her received qubit with the outcome $a$, the state changes to $\rho_{\hat{A}_x}^a=\hat{U}_{\hat{A}_x}^a.\rho.(\hat{U}_{\hat{A}_x}^a)^{\dagger}$, where $\hat{U}_{\hat{A}_x}^a=\hat{\Pi}_{\hat{A}_x}^a\otimes I, \hat{\Pi}_{\hat{A}_{x}}^a=\frac{I+a\times \hat{A}_{x}}{2}$. Then, Bob performs a weak measurement on his received another qubit with $b$, the reduced state can be given as $\rho_{\hat{B}_y}^b=\frac{F}{2}\times\rho_{\hat{A}_x}^a+\frac{1+b\times G-F}{2}\left[\hat{U}_{\hat{B}_y}^+.\rho_{\hat{A}_x}^a.(\hat{U}_{\hat{B}_y}^+)^\dagger\right]+\frac{1-b\times G-F}{2}\left[\hat{U}_{\hat{B}_y}^-.\rho_{\hat{A}_x}^a.(\hat{U}_{\hat{B}_y}^-)^\dagger\right]$, where $\hat{U}_{\hat{B}_y}^b=I\otimes\hat{\Pi}_{\hat{B}_y}^b$. $F$ is the quality factor which represents the undisturbed extent of the state of Bob’s qubit after he measured and $G$ is the precision factor which is the strength of the weak measurement and quantifies the information gained from Bob’s measurements. Subsequently, Charlie performs a strong measurement $\hat{C}_z$ with outcome $c$, the state changes to $\rho_{\hat{C}_z}^c=\hat{U}_{\hat{C}_z}^c.\rho_{\hat{B}_b}^b.(\hat{U}_{\hat{C}_z}^c)^\dagger$, where $\hat{U}_{\hat{C}_z}^c=I\otimes\hat{\Pi}_{\hat{C}_z}^c$. We can obtain the whole joint probability distribution, which is $P(a,b,c|x,y,z)=Tr\left[\rho_{\hat{C}_z}^c\right]$. According to this joint probability distribution, we can obtain the marginal probability distribution of the Alice-Bob pair, and the marginal probability distribution of the Alice-Charlie pair, which can be given as $P(a,b|x,y)=\sum_cP(a,b,c|x,y,z)$ and $P(a,c|x,z)=\sum_bP(a,b,c|x,y,z)$. Then, we can analyze the correlations of Alice-Bob and Alice-Charlie based on the marginal probability distribution. Quantum steering sharing and hybrid quantum correlation sharing are then illustrated by exploring the two linear steering inequalities and the CHSH inequality together with one linear steering inequality. 
	
	First, let's discuss the scenario where Alice steers the states of Bob and Charlie. To better observe the phenomenon of quantum steering, both measurement directions of Bob and Charlie are set vertically, i.e., $\theta_4=\frac{\pi}{2}+\theta_3, \theta_6=\frac{\pi}{2}+\theta_5$. Using the optimal solution, we can obtain the two maximal linear steering parameters, which are $S_{L_1}=\sqrt{2}G, S_{L_2}=\frac{1+F}{\sqrt{2}}$, where the optimal settings are $\theta_1=\theta_3=\theta_5=0, \theta_2=\frac{\pi}{2}$. For the weak measurements, there are two typical pointer distributions, the optimal pointer and the square pointer \cite{Silva.PhysRevLett.114.250401_2015}, where the relation between the quality factor $F$ and the precision factor $G$ satisfies $G^2+F^2=1$ or $G+F=1$, with $G, F\in\left[0,1\right]$. We employ the square pointer $G^2+F^2=1$. It is found that when $G=0.8$, the maximum value is $S_{L_1}=S_{L_2}=1.131$. 
	
	Moreover, the hybrid quantum correlation is discussed by the CHSH inequality (between Alice and Bob) and the linear steering inequality (between Alice and Charlie). Charlie's two measurement directions are set vertically, i.e., $\theta_6=\frac{\pi}{2}+\theta_5$. Using the optimal solution, we can obtain the maximal CHSH and the maximal linear steering parameter, which are $S_{C_1}=\sqrt{2}G, S_{L_2}=\frac{1+F}{\sqrt{2}}$, where the optimal setting are $\theta_1=0, \theta_2=\frac{\pi}{2}, \theta_3=\frac{\pi}{4}, \theta_4=-\frac{\pi}{4}, \theta_5=0$. Obviously, when $G=0.8$, the maximum value is $S_{C_1}=S_{L_2}=1.131$. 
	
	If Bob and Charlie aim to steer Alice's state, Alice's two measurement directions are set vertically, i.e., $\theta_2=\frac{\pi}{2}+\theta_1$. Using the optimal solution, we can obtain the two maximal linear steering parameters, which are $S_{L_1}=\sqrt{2}G, S_{L_2}=\frac{1+F}{\sqrt{2}}$, where the optimal settings are $\theta_1=\theta_3=\theta_5=0, =\theta_6=\frac{\pi}{2}$. Similarly, when $G=0.8$, the maximum value is $S_{L_1}=S_{L_2}=1.131$.
\end{document}